\documentclass{article}
\usepackage{graphicx}
\usepackage{amsmath}
\usepackage{gensymb}
\usepackage{caption}
\usepackage{siunitx}
\usepackage{subcaption}% Required for inserting images

\title{Negative thermal expansion coefficient and amorphization in defective 4H-SiC}
\author{Christopher Grome }
\date{September 2023}

\begin{document}

\maketitle

\begin{abstract}
    Silicon Carbide (SiC) is a wide bandgap semiconductor material recently being used in replacement of traditional semiconductors for high-voltage power device applications. Radiation environments induce the defects through displacement damage in the lattice that can saturate over periods of high energy particle exposure at varous concentrations, which include the formation of vacancies, interstitials and Frenkel pairs. Using molecular dynamics software we calculate thermal expansion coefficient (TEC) over and specific heat capacity at constant volume ($c_v$) values over a temperature range varying defect concentrations in single crystal 4H-SiC. At a discovered critical defect density amorphous defect clusters form in the lattice triggering macroscopic negative thermal expansion across the entire temperature range. Exponential $c_v$ loss is observed as defect density increases until the isothermal process becomes completely adiabatic at a identified critical frenkel pair concentration. Providing insight to the degradation of SiC from displacement damage effects can ultimately assist the development of radiation-hardened electronics. 
\end{abstract}

\section{Introduction}

Silicon Carbide (SiC) is a wide bandgap semiconductor material recently being used in replacement of traditional semiconductors for high-voltage power device applications \cite{roccaforte2018emerging,lauenstein2018wide,kimoto2014fundamentals}. Radiation environments that exist terrestrially and in outer space pose a noticeable risk to electronics which can lead to cumulative degradation effects within the material known as displacement damage (DD) \cite{leroy2007particle, boutte2013compendium, srour2003review}. DD characterizes the defects in the lattice that can accumulate over periods of high energy particle exposure, which include the formation of vacancies, interstitials and Frenkel pairs. Although the recombination rates for point defects in SiC are recorded to be between 46.29\% and 62.16\% for a given radiation strike \cite{chen2023simulation}, 4H-SiC has observed amorphous transformations when exposed to high enough ion fluxes \cite{ARADI}. Experimental Raman spectra analysis has shown that SiC exposed to neutron doses of .11 displacement per atom (dpa) saturate defects at an average distance of 0.6 nm yielding defect concentrations at considerable percentages \cite{koyanagi2016}. 

Displacement damage effects are have been known to cause a variety of device malfunction. Effects on device’s majority carrier density, carrier mobility, carrier lifetimes and other electronic properties have been well documented \cite{9,1,3,24,37}. However, The irradiation induced point defects affect on the macroscopic thermal and mechanical properties of 4H-SiC are not very clear.

To answer questions that are difficult to observe on-line during experiments, a theoretical study at a fundamental atomic level is employed. Molecular Dynamics (MD) is a very appropriate simulation approach for understanding material processes as it provides high fidelity studying capability of atomic-level events. MD has also been most notably employed for addressing a number of key problems in semiconductor thermal \cite{54,56}, mechanical \cite{10,57} and electrical \cite{58} properties. 

In this paper, artificially produced point defects (vacancy, interstitial and Frenkel pairs) are stochastically generated in bulk 4H-SiC. The thermal expansion coefficient (TEC) and specific heat capacity ($c_{v}$) are investigated over a temperature range of 200K to 1200K.The effects of defect type and concentration are quantified and an advancement in the understanding of the thermal mechanics of defected systems is made.

Studying the TEC is important for semiconductor materials. Compatible TECs are required between substrates and at bonded material interfaces to minimize thermal stress and optimize device performance \cite{44}. Variation in the TEC can result in incompatible thermal stresses that can cause permanent damage in semiconductor devices \cite{45,46}. Specific heat capacity is an imperative parameter for conductance and thermal management of semiconductor materials \cite{48}. Deviations in the specific heat can lead to premature changes in the temperature of power devices consequently effecting all temperature dependent electronic parameters leading to performance degradation. Additionally, specific heat has been shown to effect the susceptibility of devices to radiation effects. Simulation work in \cite{27,49,50}, shows heat capacity critically effects thermal modulation in power devices during single event effects (SEE). Changes to this parameter leave power electronics more prone to reaching its sublimation point during high voltage stress or radiation strikes, resulting in permanent degradation or catastrophic failure.

 The goal and motivation behind this research is to provide insight to the degradation of the material in extreme environments to ultimately assist the development of radiation-hardened electronics with applications in nuclear reactor monitoring, aerospace, and deep space exploration.

\begin{figure}
     \centering
     \begin{subfigure}[b]{0.3\textwidth}
         \centering
         \includegraphics[width=\textwidth]{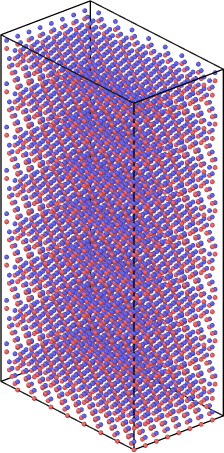}
         \caption{Pure 4H-SiC supercell}
         \label{fig:puremodel}
     \end{subfigure}
     \hfill
     \begin{subfigure}[b]{0.3\textwidth}
         \centering
         \includegraphics[width=\textwidth]{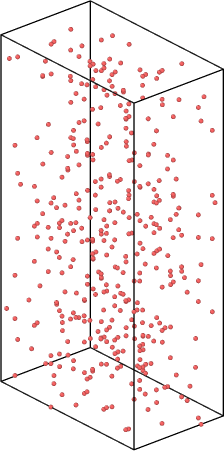}
         \caption{with vacancies}
         \label{fig:vacancies}
     \end{subfigure}
     \hfill
     \begin{subfigure}[b]{0.3\textwidth}
         \centering
         \includegraphics[width=\textwidth]{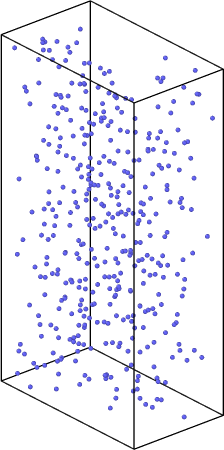}
         \caption{with interstitials }
         \label{fig:interstitials}
     \end{subfigure}
     \hfill
     \begin{subfigure}[b]{0.3\textwidth}
         \centering
         \includegraphics[width=\textwidth]{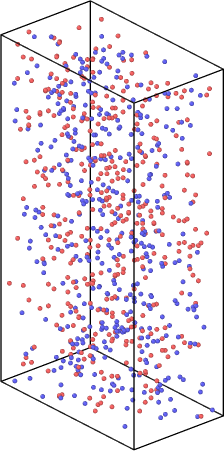}
         \caption{with Frenkel Pairs}
         \label{fig:frenkels}
     \end{subfigure}
        \caption{Modeling of 4H-SiC supercells a) pure system b) vacancy defects in red c) Interstitial defects in blue d) Frenkel pairs}
        \label{fig:defectmodels}
\end{figure}

\section{Computational Methods}

Modeling the displacement damage defects on single crystal 4H-SiC and its effect on the materials thermal and mechanical properties requires the use of a toolkit and a simulator. Displacement damage and non-ionizing energy loss mechanisms are inherently atomistic-scale events. To comprehensively map the physical movements of atoms and observe the evolution of the system after DD defects are imposed, a molecular dynamics (MD) simulator was employed. A compatible toolkit was developed to provide the MD software with atomic models of defected 4H-SiC. 

The MD simulations carried out in this work are with the software LAMMPS (Large-scale Atomic/Molecular Massively Parallel Simulator). The Modified Embedded Atom Method (MEAM) potential is the first semi-empirical interatomic potential formalism that demonstrated the possibility that a single formalism can be applied to a wide range of structured elements \cite{59}. The MEAM potential proposed by Kang \emph{et al.} is used in this work to describe inter-atomic and point defect interactions.  For modeling the body interactions, LAMMPS uses the notation Si-C. The MEAM library and SiC parameter files used for MD are taken from \cite{61,62} and can be found in the Appendix. Using the .dat file generated from the modeling toolkit, LAMMPS can compute thermal simulations of interest. 

\subsection{4H-SiC Models}
The super cell of pure 4H-SiC is generated with dimension 8a $\times$ 8b $\times$ 8c consists of 8192 atoms, developed using known lattice constants, dimensional properties given in Table \ref{table:1} and stacking sequence ABCB. This structure has a space group 186 (P 63 m c) and point group 6mm ($C_{6V}$) one 6 fold rotation, six mirror planes and no inversion \cite{51,52}. 

\begin{table}[h!]
\centering
\begin{tabular}{||c c ||} 
 \hline
 Lattice Constant & Dimension \\ [0.5ex] 
 \hline\hline
 a & 3.079 \AA \\ 
 b & 3.079 $\sqrt{3}$ \AA \\
 c &  10.053 \AA \\
 $\alpha$ & 90\degree  \\
 $\beta$ & 90\degree \\ 
  $\gamma$ & 120\degree \\ [1ex] 
 \hline
\end{tabular}
\caption{Lattice constants for 4H-SiC unit cell}
\label{table:1}
\end{table}

It has been documented that defects from displacement damage are inherently stochastic in nature \cite{1,24,10}. Ergo, our modeling techniques introduce point defects into the system randomly and uniformly. For vacancy defects atoms are randomly removed, interstitial defects randomly add atoms to the system and Frenkel pairs randomly add and remove atoms in the system. In this work, the defects are introduced separately and uniformly at concentrations of 2\%, 4\%, 6\%, 8\% and 10\% (as percentages of the total atom count) to investigate the role the defect type and density have on the materials properties. The defected models used in this with their respective atom counts are listed in Table \ref{table:2} and are visualized in Figure \ref{fig:defectmodels}

\begin{table}[h!]
\centering
\begin{tabular}{||c c c c c c ||} 
 \hline
 Defect Density: & 2\% &  4\% &6\% &8\% & 10\%\\ [0.5ex] 
 \hline\hline
 Vacancy Models (\# of atoms) & 8028 & 7864 & 7701 & 7537 & 7373 \\ 
 Interstitial Models (\# of atoms) & 8356 & 8520 & 8683 & 8847 & 9011 \\ 
 Frenkel Pair Models (\# of atoms) & 8192 &8192 & 8192& 8192 & 8192 \\  [1ex] 
 \hline
\end{tabular}
\caption{Atomic counts for defected models}
\label{table:2}
\end{table}

\subsection{Thermal Simulation Modeling}
The energy data of these samples is calculated with LAMMPS to study the effect on the thermal properties. In these simulations, we initially relax the system by applying a conjugate gradient minimizer to dissipate residual stress. Following, we assign random velocity distributions to the atoms to represent 1K. We then apply a Noose-Hoover thermostat to raise the system to the target temperature over a period of 5ns with a time step of 1 fs. The pressure damping coefficient in the NPT ensemble simulation were tailored for each temperature simulation to converge to the experimental and literature data. A breakdown of the damping coefficient parameters is provided in Table \ref{table:3}. Periodic boundary conditions are applied along the [111] direction and the supercells orthogonal sides are bounded by the planes with normal directions along the [111], $[1\,\overline{1}\,0]$ and $[1\,1\,\overline{2}]$ directions. For structure analysis, Open Visualization Tool (OVITO) is employed. 

\begin{table}[h!]
\centering
\begin{tabular}{||c c ||} 
 \hline
 Temperature (K) & Pdamp (fs) \\ [0.5ex] 
 \hline\hline
 200 & 320.0 \\ 
 300 & 245.0 \\
 400 &  199.3 \\
600 & 285.9  \\
 800 & 250.0 \\ 
  1000& 490.0 \\
   1200& 396.0 \\ [1ex] 
 \hline
\end{tabular}
\caption{Pressure damping coefficients (Pdamp) for TEC conversion at different temperatures }
\label{table:3}
\end{table}

\section{Thermal Simulation Results}

\begin{figure}[htp]
    \centering
    \includegraphics[scale=0.40]{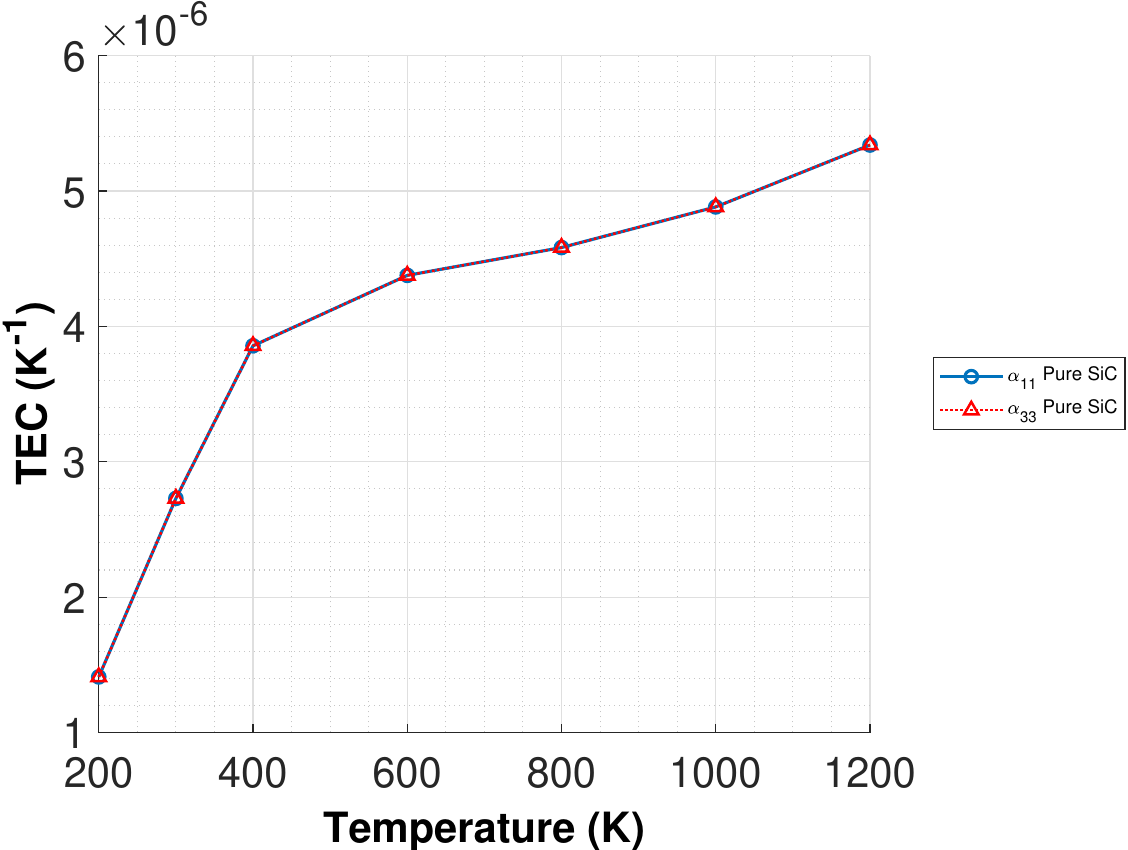}
    \caption{Directional dependence of TECs calculated along a and c axes}
    \label{fig:isofig}
\end{figure}

A number of papers have been reported on the TEC of SiC. Work in \cite{43} showed anisotropic dependence, where the a-axis TEC was larger than the TEC measured on the c-axis. However in more recent study using both laser interferometry and dilatometry techniques, Nakabayashi et al. shows there is an isotropic relationship. A review in \cite{47} recorded more than 30 published thermal expansion measurements had no reported anisotropy for various silicon carbide materials. Our directional dependence study in Figure \ref{fig:isofig} shows our MD simulations also are isotropic. The TEC values between the c-axis ($\alpha_{33}$) direction and the a-axis ($\alpha_{11}$) direction were exactly the same to six decimal places.

\begin{figure}
     \centering
     \begin{subfigure}[b]{0.45\textwidth}
         \centering
         \includegraphics[width=\textwidth]{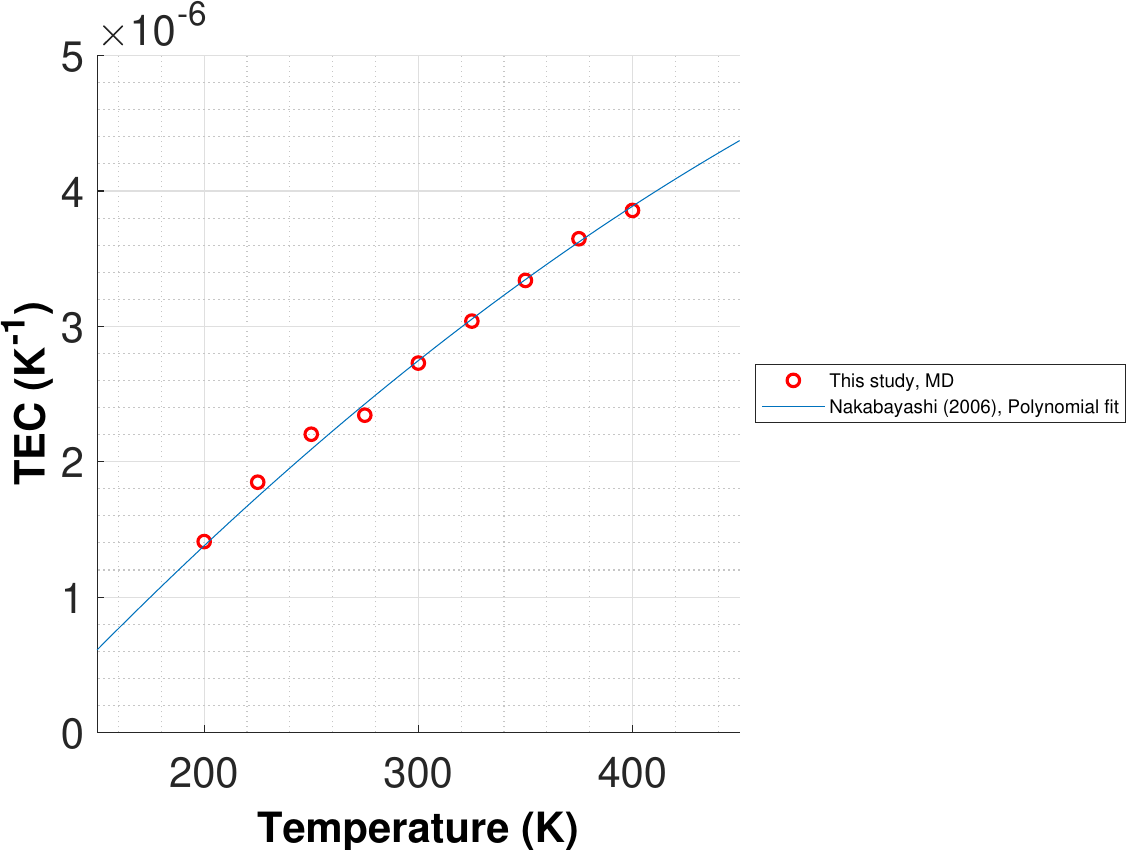}
         \caption{}
         \label{fig:lowtemp}
     \end{subfigure}
     \hfill
     \begin{subfigure}[b]{0.45\textwidth}
         \centering
         \includegraphics[width=\textwidth]{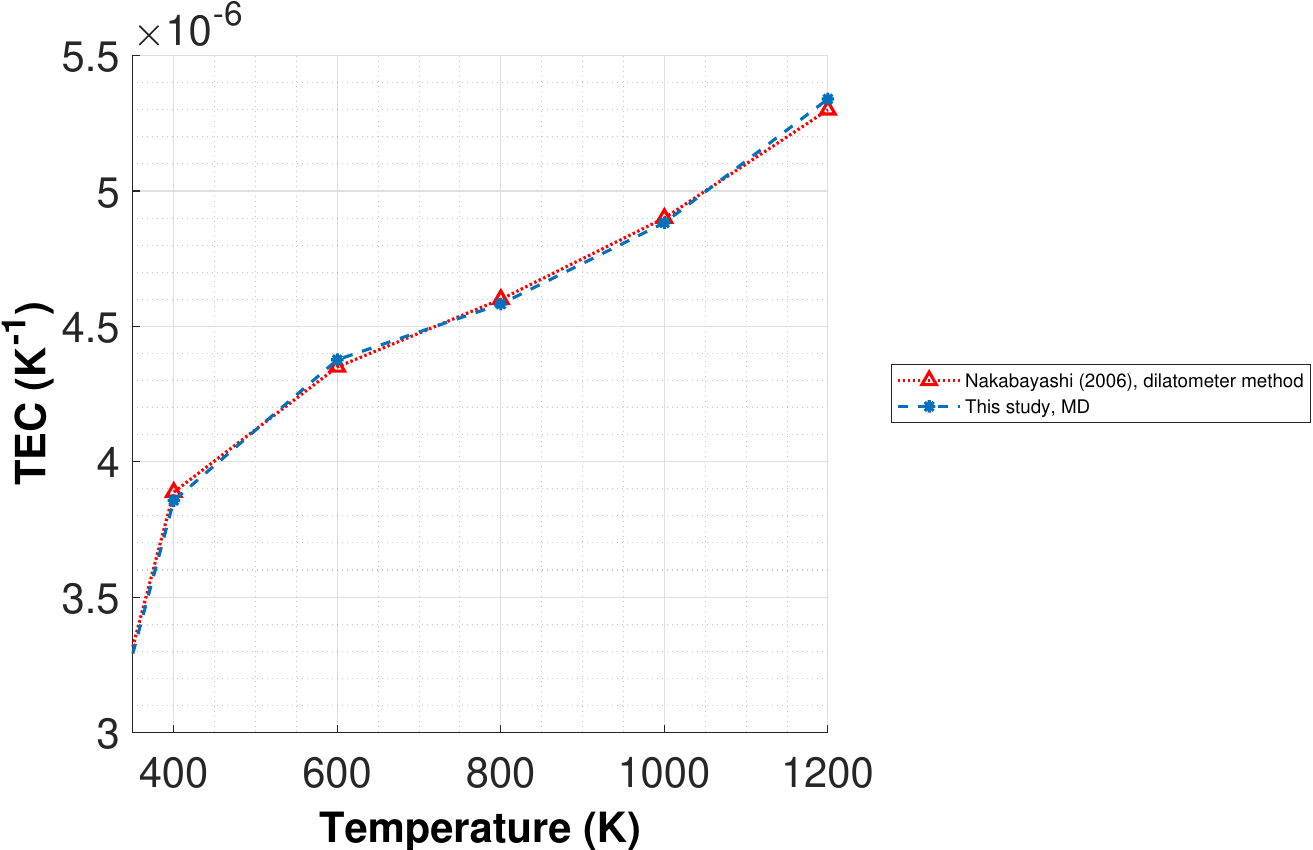}
         \caption{}
         \label{fig:hitemp}
     \end{subfigure}
     \hfill
      \caption{a) Linear Thermal Expansion Coefficient for low temperatures comparing MD simulations to analytical polynomial b) LTEC for high temperatures compared to experimental data}
    \label{fig:TECvalidation}
\end{figure}

The linear thermal expansion coefficients were taken along the $\langle$11-20$\rangle$ direction and were dynamically compared at both low and high temperatures to literature. Using the ensemble bond length at equilibrium, we calculate the temperature dependent linear thermal expansion coefficient ($\alpha$) along the a-axis from Equation \ref{eqn:TEC}.

\begin{equation}
\label{eqn:TEC}
    \alpha=\biggl(\frac{dln(L)}{dT}\biggr)_{p}
\end{equation}
The low temperatures were ran every 25K from 200K to 400K and were compared to an adopted analytical TEC polynomial from \cite{6} that fit the experimental data from 123K to 473K in their work. The polynomial for the TEC along the a-axis is 
\begin{equation}
\label{eqn:low t poly}
\alpha_{11} = -2.0404+1.9374 \times10^{-8}\cdot T +1.1385\times 10^{-11}\cdot T^2  (K^{-1})
\end{equation}
and is plotted in comparison to the results from our MD in Figure \ref{fig:lowtemp}. As illustrated our MD shows very good agreement with the literature results and can attribute fidelity to our simulation. At higher temperatures, our MD values are compared to individual experimental values every 200K from 400K to 1200K in Figure \ref{fig:lowtemp}.

Our MD simulation agreement with analytical and experimental results indicates the second nearest neighbor distance dominates the LTEC behavior due to the MEAM potentials inherent nearest-neighbor model. The results for the pure structure show positive and values and a square root relationship to temperature. As temperature increases the potential energy decreases which represents the stiffening and softening of the state of the bond. The kinetic energy of the system is increased and the bond lengths between the atoms expand. At high temperatures, our simulation continued to show good agreement with literature and allows for defect analysis.		
Specific heat capacity for SiC has also been well studied and is very useful for quantifying internal energy changes within a system. Work in \cite{47} shows the change in 4H-SiC $c_{v}$ as temperature increases is due to more phonon modes being made available. As these modes become available and occupied, the internal energy increases at a faster rate. Ergo, the lattice structure needs more energy to increase the temperature by a degree in higher temperature environments. An analytical 4H-SiC $c_{v}$ model was developed in \cite{47} from experimental measurements of the specific heat up to its sublimation temperature and is used to validate our MD results at T=300K. 

\begin{equation}
c_{v}=\biggl(\frac{dE}{dT}\biggr)_{V} \label{eqn:Cv}
\end{equation}

The TEC MD simulations gives us the internal energy data needed to compute $c_{v}$. Using Equation \ref{eqn:Cv}, our MD at T = 300K gives us a $c_{v}$ of 2.568 \unit[inter-unit-product = \ensuremath{{}\cdot{}}]{\joule\per\kelvin\per\centi\metre\cubed}. Using the polynomial developed in [47] at T = 300K, the analytical $c_{v}$ is 2.628 \unit[inter-unit-product = \ensuremath{{}\cdot{}}]{\joule\per\kelvin\per\centi\metre\cubed}, showing good agreement with our simulation. Using our simulation value as a baseline, the analysis explored will be looking at the $c_{v}$ dependency on defect density at T = 300K. 

\subsection{TEC of Defected 4H-SiC}
Linear thermal expansion coefficient curves for 4H-SiC supercells with vacancies, interstitials and Frenkel pairs at varied concentrations are calculated at 7 different temperatures ranging from 200 to 1200K. Using the bond length at equilibrium, we calculate TEC along the a-axis using Equation \ref{eqn:TEC}. We consider defect concentrations from 2 to 10\% and observe the LTEC defect density dependence. Due to the stochastic algorithm used to employ We model 5 iterations of each defect concentration

\begin{figure}
     \centering
     \begin{subfigure}[b]{0.45\textwidth}
            \centering
            \includegraphics[scale=0.40]{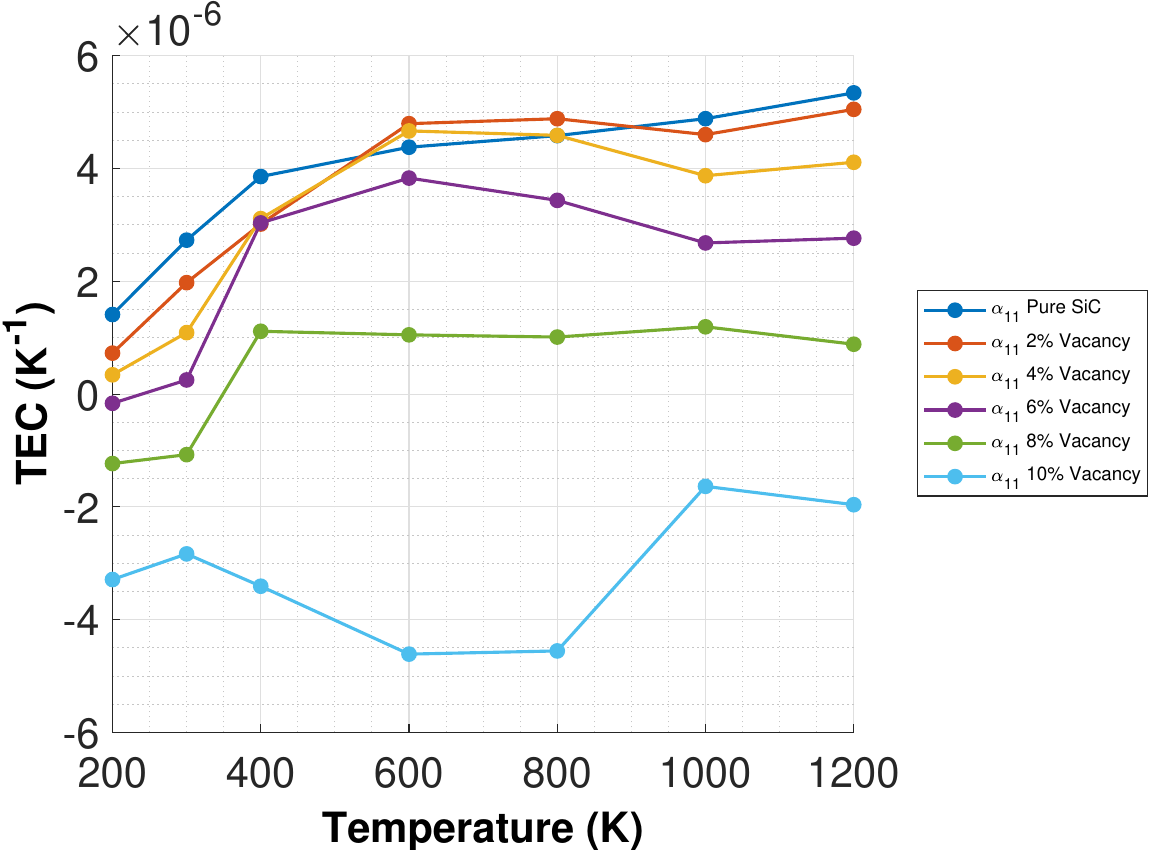}
            \caption{Vacancies}
            \label{fig:vactecfig}
     \end{subfigure}
     \hfill
     \begin{subfigure} [b]{0.45\textwidth}
            \centering
            \includegraphics[scale=0.40]{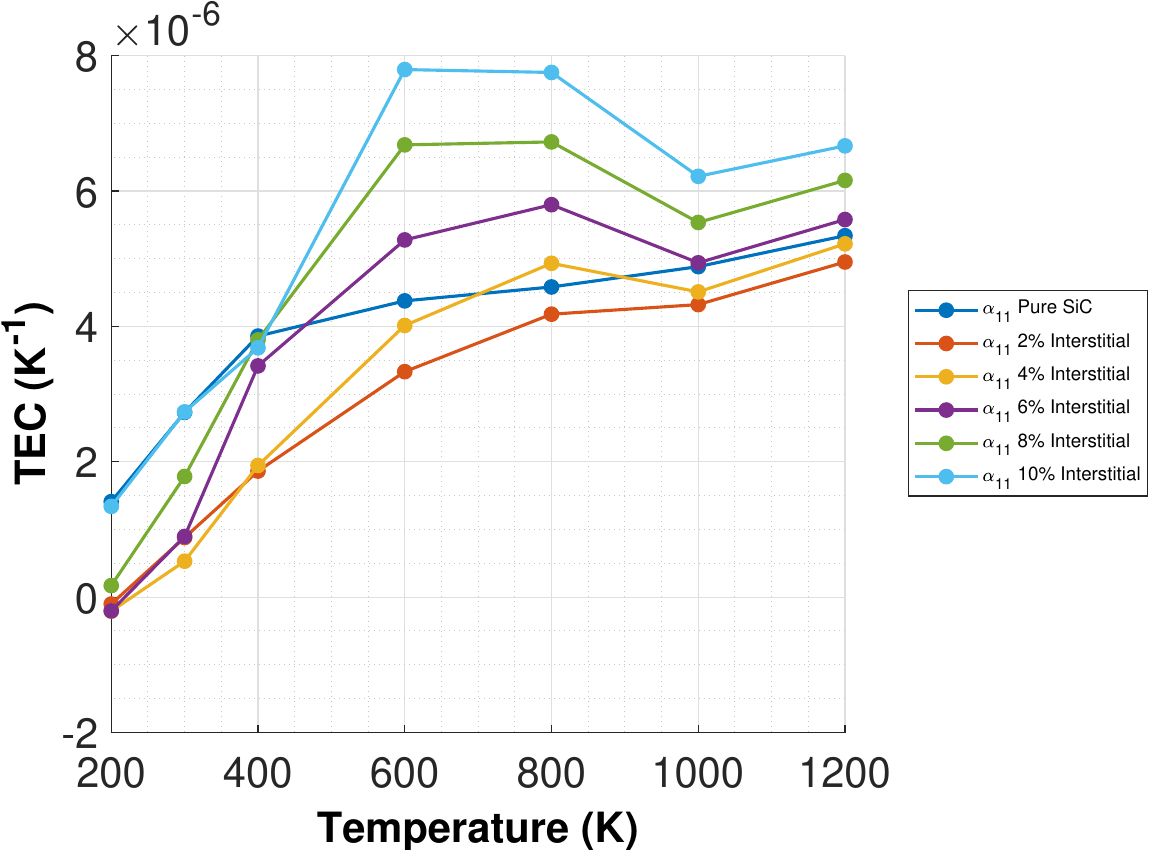}
            \caption{Interstitials}
            \label{fig:inttecfig}
     \end{subfigure}
     \hfill
     \begin{subfigure}[b]{0.45\textwidth}
            \centering
            \includegraphics[scale=0.40]{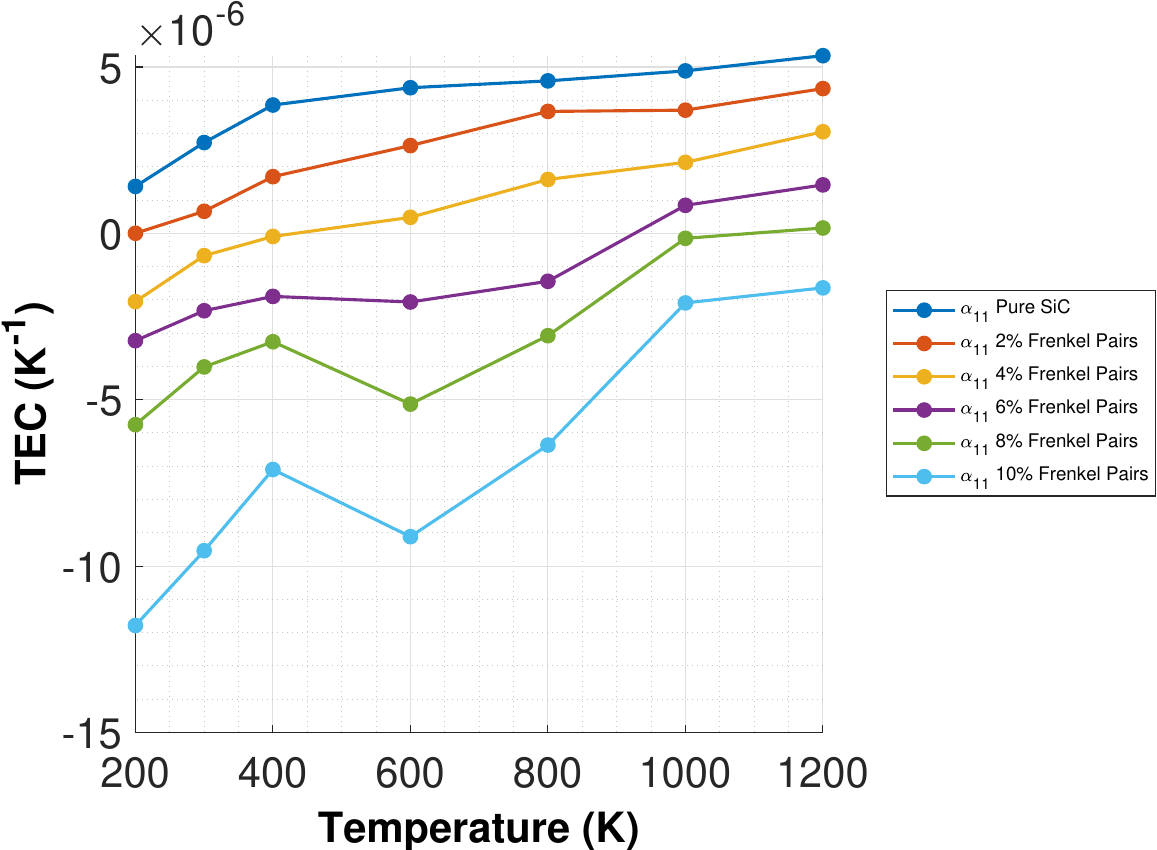}
            \caption{Frenkel Pairs}
            \label{fig:frenktecfig}
     \end{subfigure}
     \hfill
        \caption{TEC of (a) vacancy (b) interstitial and (c) defected 4H-SiC at various concentrations in comparison to the non-defected material}
        \label{fig:TEC}
\end{figure}

Observations on the effect of vacancy defects are first discussed. The $\alpha_{11}$ for all vacancy defect densities ($\rho_{vd}$) are obtained and compared to the baseline pure SiC. The results presented in Figure \ref{fig:vactecfig} suggest a number of important features in vacancy defected 4H-SiC: (1) $\alpha_{11}$ is not highly vacancy density dependent until $\rho_{vd} > 6\% $; (2) for $\rho_{vd}$ $\leq{6\%}$ the square root temperature dependence is followed until higher temperatures T $>$ 600K, for 800K$<$T$<$ 1200K a decline in $\alpha_{11}$ is observed; (3) for $\rho_{vd}$ = 8\% a negative $\alpha_{11}$ is observed during T $<$ 400K, then showing $\alpha_{11}$ temperature independence for T $\geq$ 400K; (5) there exist a critical vacancy defect value 8\% $<$ $\rho_{vd}$ $<$ 10\% that triggers a complete thermal contraction in the material. 

The TEC data at various interstitial defect densities $\rho_{id}$ provided in Figure \ref{fig:inttecfig} suggests several important features in interstitial defected 4H-SiC: (1) $\alpha_{11}$ is not highly interstitial density dependent until $\rho_{id} \> 6\%$; (2) for $\rho_{id} < 6\%$ the systems under-expand the baseline case; (3) for $\rho_{id} \geq 6\%$ the systems over expand the baseline case; (4) for $\rho_{id} < 8\%$, increases in temperature lead to closer agreement with the pure SiC $\alpha_{11}$; (5) interstitials trigger an inherent over-expansion phenomena in the system that increases in magnitude with respect to $ \rho_{id} $.

Lastly, observations on the effect of Frenkel pairs are discussed. Figure \ref{fig:frenktecfig} suggests: (1) the largest TEC deviations per defect density are caused by frenkel pairs; (2) increasing $\rho_{fp}$ exclusively underexpands the baseline case; (3) square root temperature dependence seen in  $\rho_{fp} \leq 4\%$, uniform dip at 600K in remaining defected cases; (4) for $T \leq 400K$ negative thermal expansion (NTE) is observed in all, but  $\rho_{fp}$ = 2\% case; (5) exclusive NTE observed for $\rho_{fp}$ = 10\% case.

\begin{figure}
     \centering
     \begin{subfigure}[b]{0.45\textwidth}
         \centering
         \includegraphics[width=\textwidth]{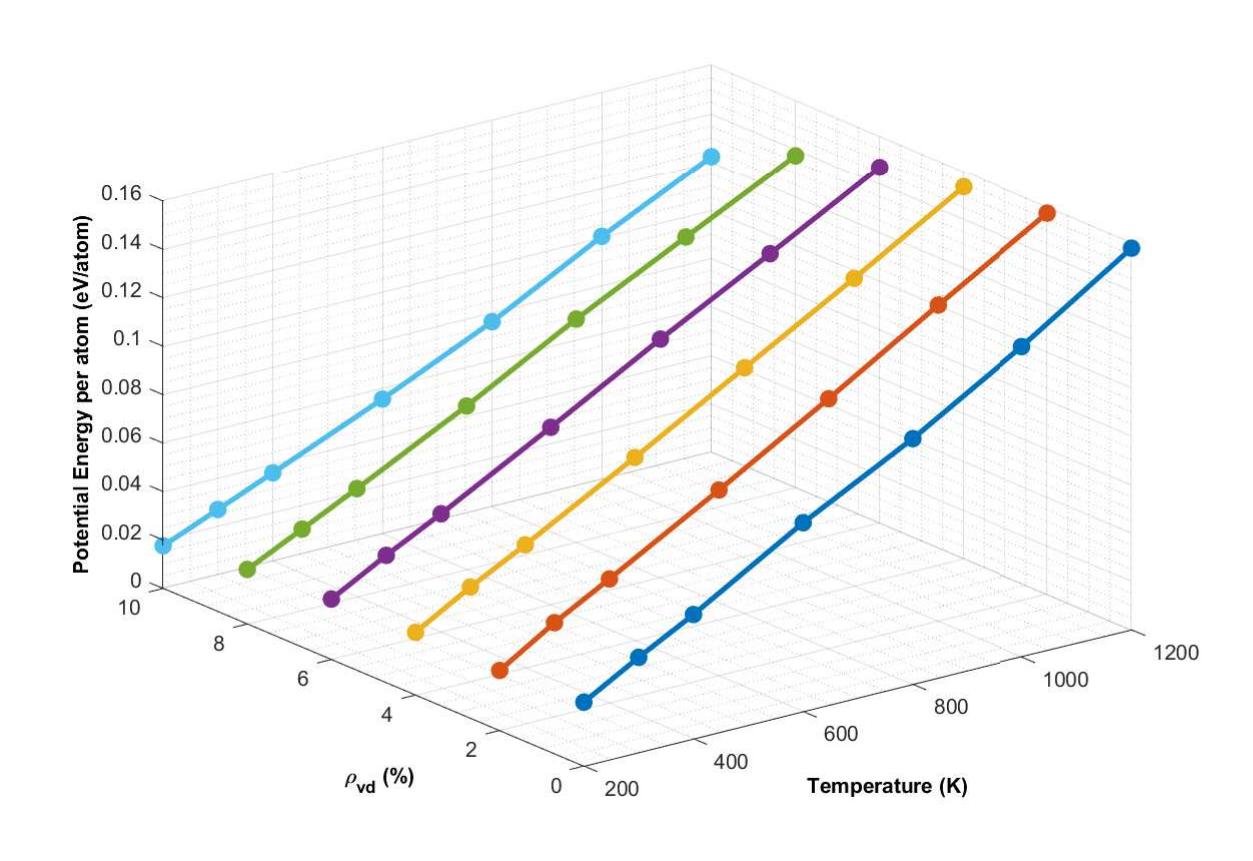}
         \caption{}
         \label{fig:PEvacandpure}
     \end{subfigure}
     \hfill
     \begin{subfigure}[b]{0.45\textwidth}
         \centering
         \includegraphics[width=\textwidth]{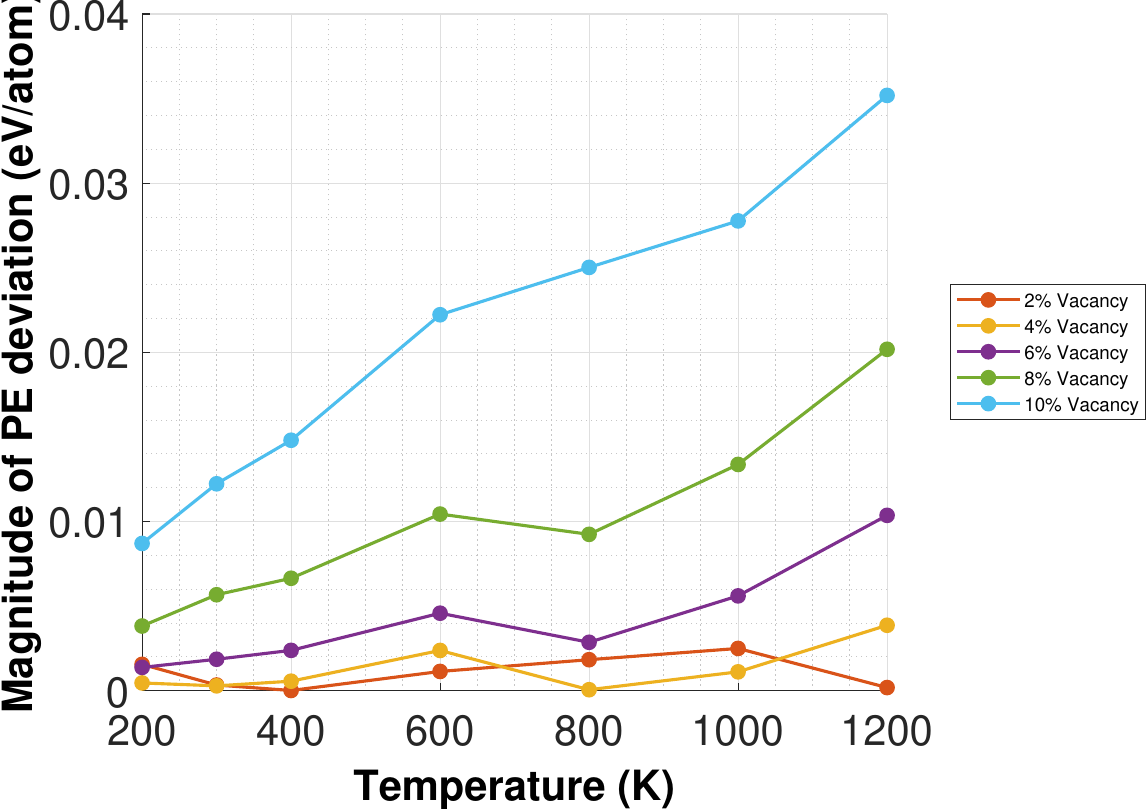}
         \caption{}
         \label{fig:diffPEvac}
     \end{subfigure}
     \hfill
          \begin{subfigure}[b]{0.45\textwidth}
         \centering
         \includegraphics[width=\textwidth]{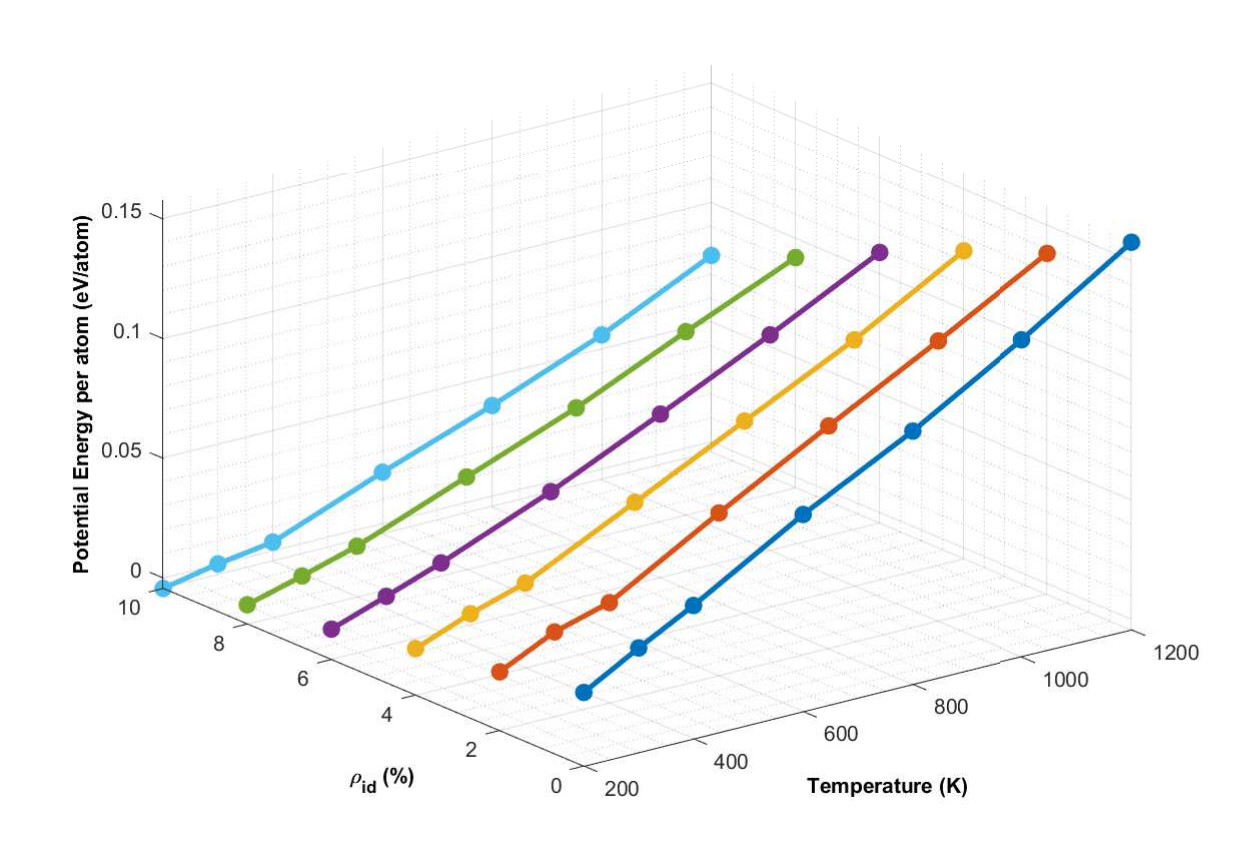}
         \caption{}
         \label{fig:PEintandpure}
     \end{subfigure}
     \hfill
     \begin{subfigure}[b]{0.45\textwidth}
         \centering
         \includegraphics[width=\textwidth]{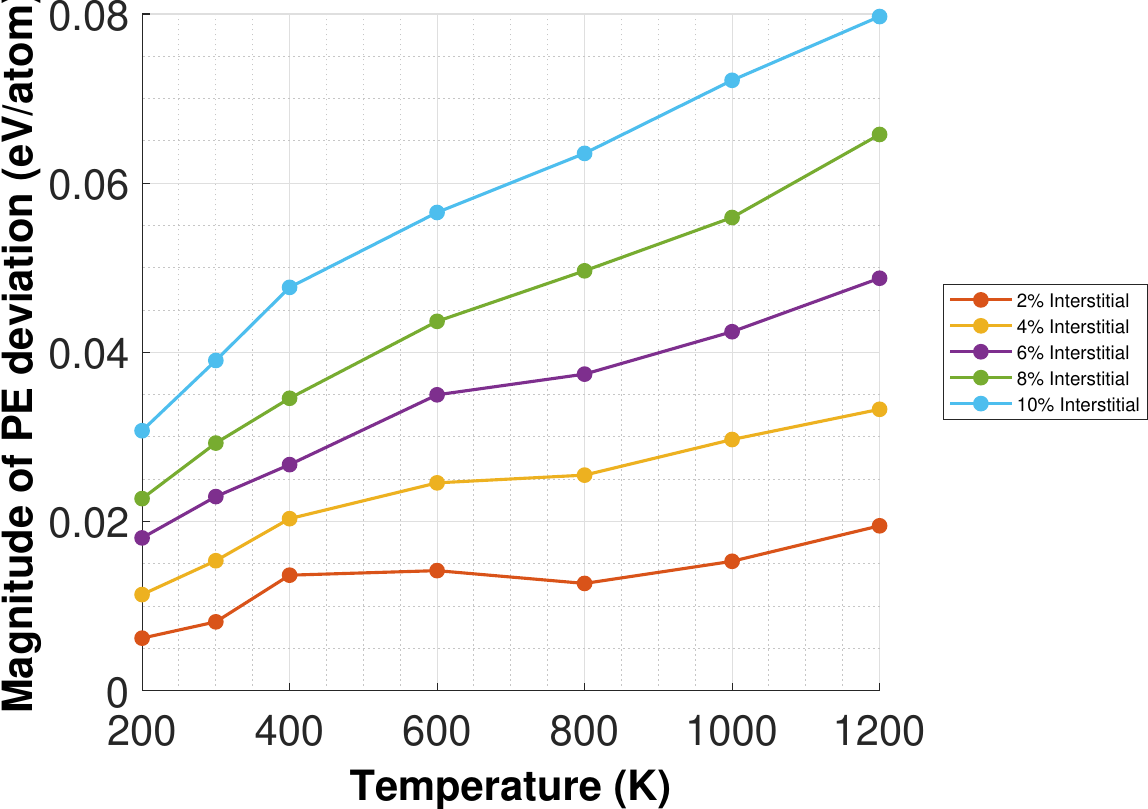}
         \caption{}
         \label{fig:diffPEint}
     \end{subfigure}
     \hfill
          \begin{subfigure}[b]{0.45\textwidth}
         \centering
         \includegraphics[width=\textwidth]{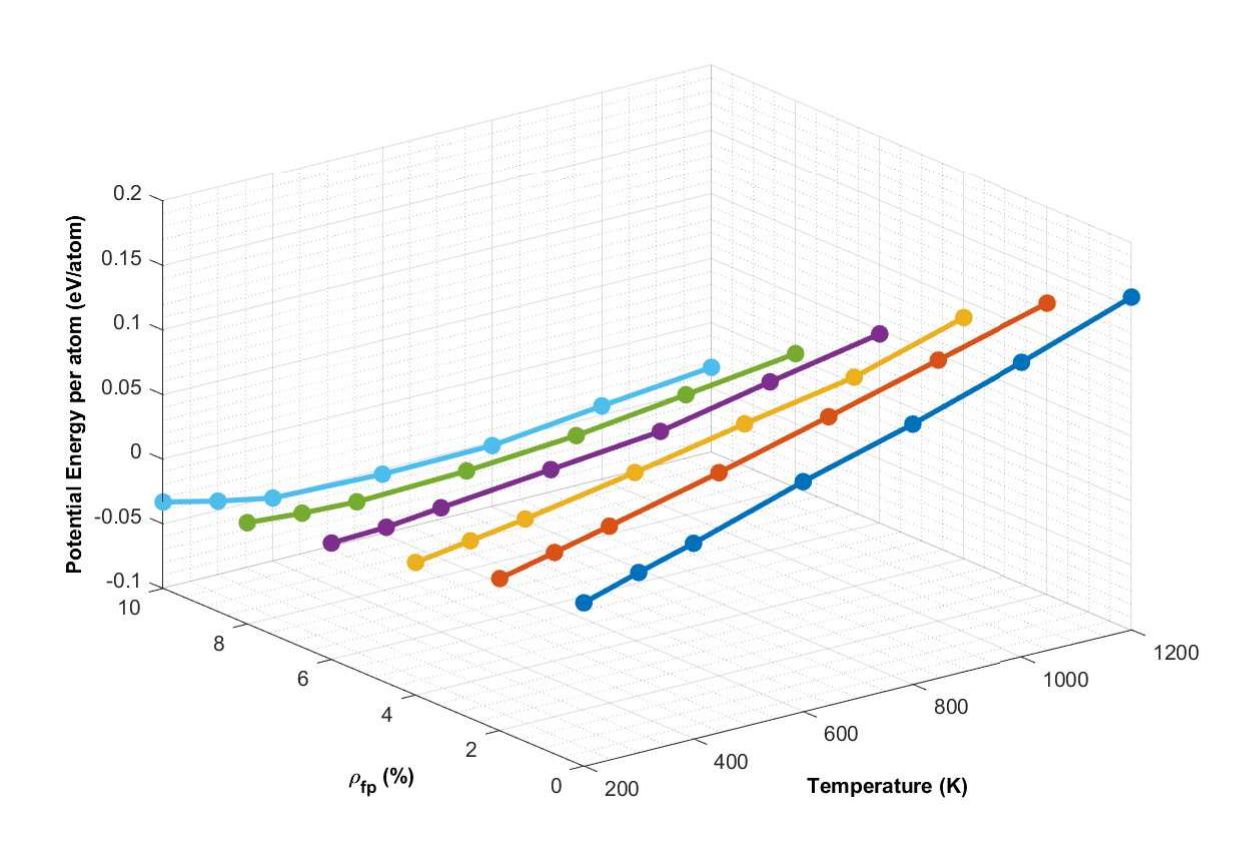}
         \caption{}
         \label{fig:PEfpandpure}
     \end{subfigure}
     \hfill
     \begin{subfigure}[b]{0.45\textwidth}
         \centering
         \includegraphics[width=\textwidth]{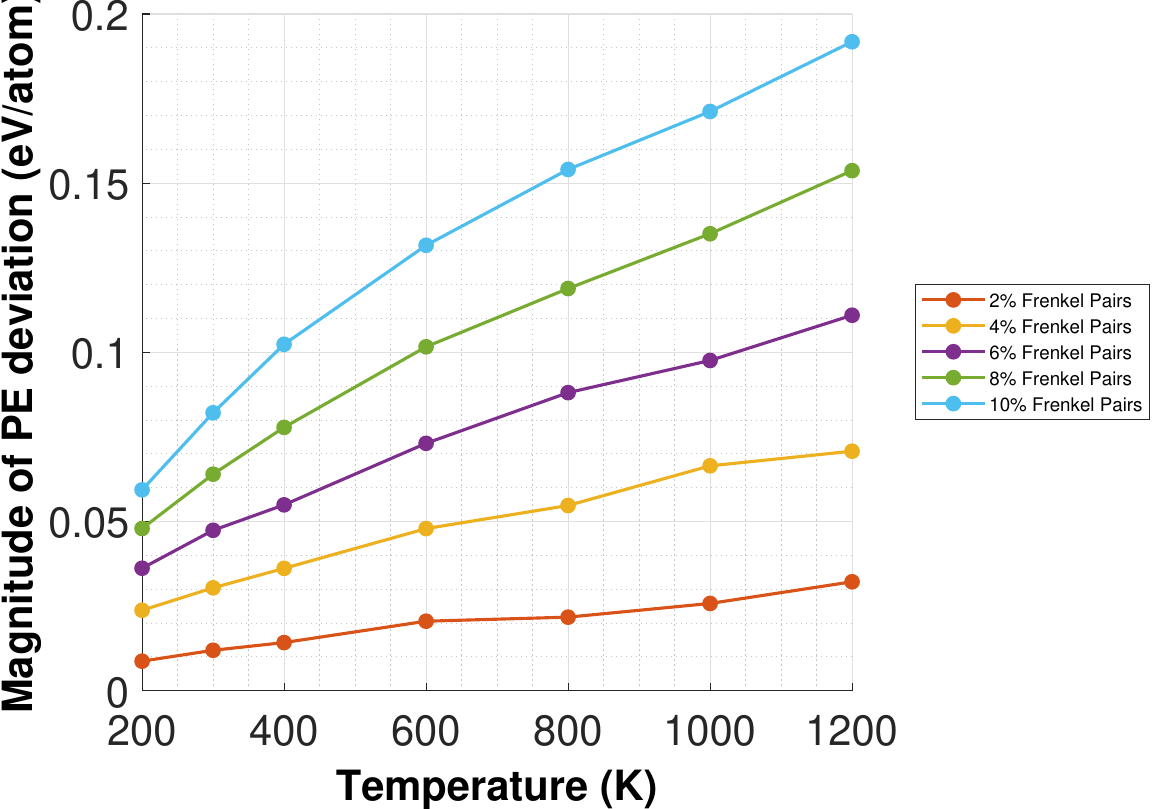}
         \caption{}
         \label{fig:diffPEfp}
     \end{subfigure}
     \hfill
      \caption{a) Vacancy, (c) interstitial and (e) Frenkel pair defect density dependence on change in systems potential energy per atom during thermal expansion. Comparison of (b) vacancy, (d) interstital and (e) Frenkel pair of structures magnitude of deviation in potential energy (PE) per atom from the baseline case }
    \label{fig:PEdef}
\end{figure}

\subsection{$C_v$ of Defected 4H-SiC}

\begin{figure}[htp]
    \centering
    \includegraphics[scale=0.40]{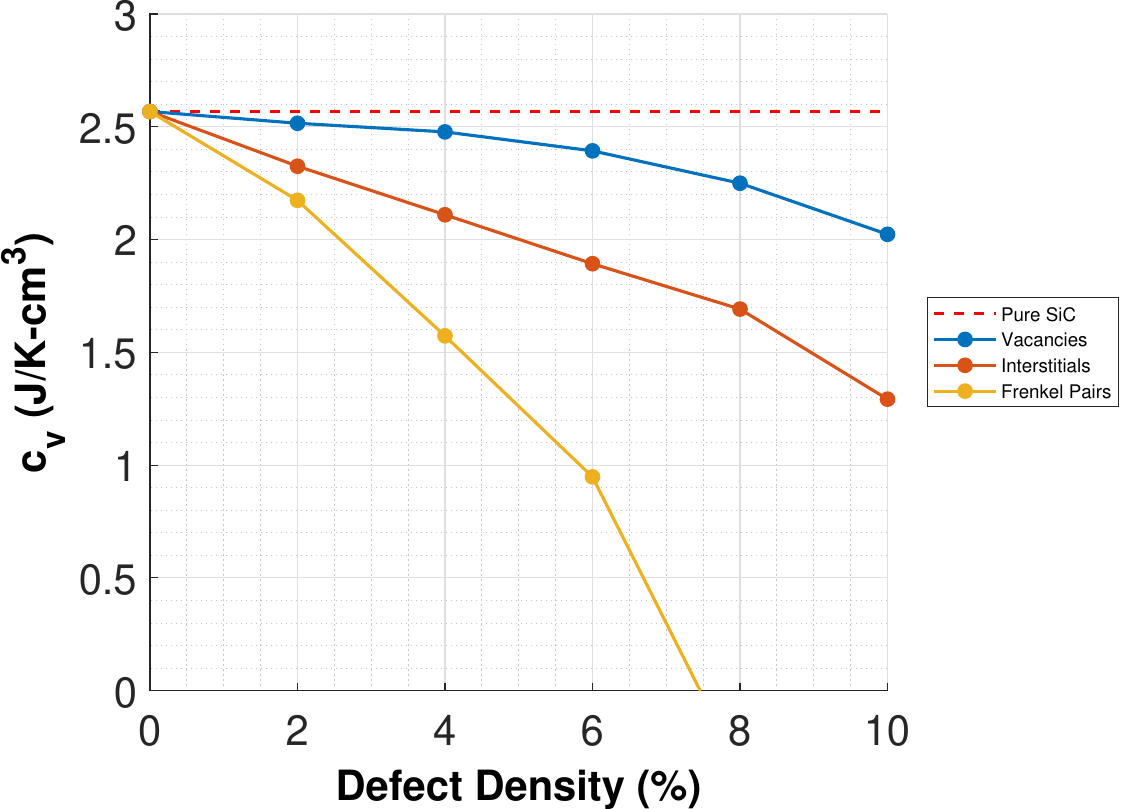}
    \caption{Specific heat capacity dependency on defect type and defect density at 300K.}
    \label{fig:cvfig}
\end{figure}

To quantify the effect defects have on the energy required to raise the temperature on the material, the specific heat capacity at constant volume is calculated in comparison to the baseline case at 300K. Figure \ref{fig:cvfig} suggests some interesting characteristics: (1) $c_v$ decreases in an exponential fashion with increased defect density; (2) Frenkel pairs have the highest decreasing rate amongst the defect types, dropping 63\% by $\rho_{fp}$ = 6\%; (3) vacancies show the least rapid decrease in energy modulation, dropping 21.2\% of the baseline when $\rho_{vd}$ = 10\%; (4) Frenkel pairs have a critical defect value between 6\% and 8\% that causes the system to not require energy to raise the temperature of the material. The specific heat capacity for $\rho_{fp} \geq 8\%$ suggests an amorphous effect compromises the materials thermal stability. 

Extrapolating, there exists critical concentrations for all defect types that will compromise the enthalpy yielding no material stability at moderate temperatures. It is acknowledged that reaching a defect concentration for interstitials and vacancies alone to achieve that level of criticality would require such extreme environment exposure that would not be within the scope of electronic applications. Overall, the data suggests that defected systems do not behave like typical lattice solids. As defects increase in concentration less energy is required to raise the temperature of the material. This is a very important parameter in radiation hardening, as defected systems with significantly impacted heat capacity will be much more susceptible to reaching melting/sublimation temperatures and premature burnout. 

\subsection{Dynamics of Defected 4H-SiC}

The dependency of $\alpha_{11}$ on defect density can be explained by exploring the potential energy (PE) of the system during thermal expansion. As temperature is introduced into the statically equilibrated pure 4H-SiC structure, the potential energy increases, but since PE is inherently a negative value the magnitude decreases. Note that the change in the potential energy during thermal expansion is the equivalent of the kinetic energy added into the system. The PE of the system is calculated by summing the bond energies within the system, so when adding and removing atoms in the system by introducing defects, changes in PE are anticipated. In order to quantify the effect the defects have on the material, the system energy data is normalized by calculating the PE per atom. The relationship of the potential energy per atom during thermal expansion for defected systems is displayed in Figure \ref{fig:PEvacandpure},\ref{fig:PEintandpure} and \ref{fig:PEfpandpure}. To elucidate the relationship between defect density and PE deviation from the baseline, the magnitude of the difference in defected potential energies to the pure 4H-SiC are taken in Figures \ref{fig:diffPEvac}, \ref{fig:diffPEint} and \ref{fig:diffPEfp}. Figure \ref{fig:PEdef} provides some additional features: (1) magnitude of PE deviation for frenkel pair concentrations are an order of magnitude higher than the other defect types; (2) the rate of PE deviation and TEC deviation increases at an increasing rate with respect to $\rho_{vd}$, rate of increase for other defect types are uniform; (3) deviation in PE from the baseline is a function of temperature; (4) Figure \ref{fig:PEfpandpure} shows for $T \leq 1000K$ PE per atom for $\rho_{fp} \geq 8\%$ is negative during thermal expansion, showing largest magnitude at T = 600K. 

The potential energies are calculated by the pair and bond energies between atoms which directly affects the enthalpy (H) of the system. In turn, these potentials provide internal energy contributions to the Gibbs free energy (G) equation in Equation \ref{eqn:Gibbs}, which describes the thermodynamic process to reach system equilibration as a function of internal energy (U), pressure (p), temperature (T), volume (V) and entropy (S) after point defect formation. 

\begin{equation}
\label{eqn:Gibbs}
    \Delta{G} = \Delta{H} - T\Delta{S} = \Delta{U} + p\Delta{V} - T\Delta{S}
\end{equation}

As the enthalpy changes within the system, the volumetric response does as well. The deviation observed in $\alpha_{11}$ with respect to defect density is caused by the first law of thermodynamics. At constant temperature and constant pressure, the thermal-dynamic potential represented in Gibbs free energy changes causing variation in the volumetric expansion, directly effecting $\alpha_{11}$. When $\rho_{vd}$ $\geq 8\%$ and $\rho_{fp}$ $\geq 6\%$, the change in energy per atom surpasses a threshold in the thermal dynamic potential causing a contracting effect on the system yielding negative thermal expansion (NTE). Rather than expanding with increased temperatures, these defected systems expand with decreased temperatures. The increase of deviation in potential energy per atom suggests that increased vacancy concentrations change the anaharmonicity associated with the interatomic distances. The vibrational properties in the $\rho_{vd} \geq 8\%$ cases are quite different. The interatomic potentials yield transverse motion perpendicular to the directions of the atomic chains, creating a shortening phenomenon. The transverse phonon amplitudes outweigh the expansion effects of the longitudinal modes resulting in systematic NTE. Additionally, the nonlinear rate of increased PE and TEC deviation suggests that the defects are interacting with each other at higher concentrations to trigger more rapid thermal degeneration of the material.

In the positive thermal expansion (PTE) cases, the interatomic potential is not disrupted enough to have a the contracting effect. The anaharmonicity leads to an increase in average interatomic distances as higher vibrational states become more populated as temperature rises. The atom chains move in longitudinal vibrations in the directions of the bonds causing a lengthening effect during increased temperature.

\begin{figure}
     \centering
     \begin{subfigure}[b]{0.4\textwidth}
         \centering
         \includegraphics[width=\textwidth]{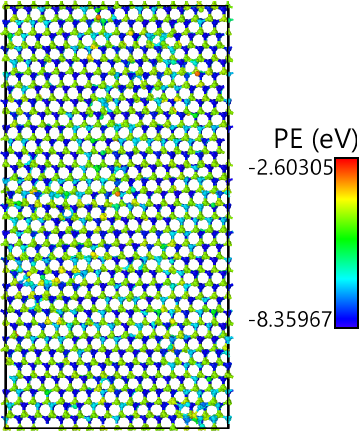}
         \caption{1K}
         \label{fig:10vacin}
     \end{subfigure}
     \hfill
     \begin{subfigure}[b]{0.4\textwidth}
         \centering
         \includegraphics[width=\textwidth]{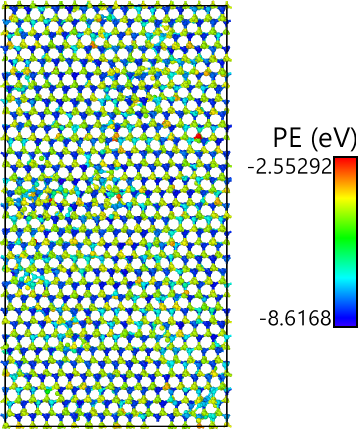}
         \caption{800K}
         \label{fig:10vacfinal}
     \end{subfigure}
     \hfill
      \caption{a) Potential Energy contour of 10\% vacancy defected model at 1K  b) Potential Energy contour of 10\% vacancy defected model at 800K }
    \label{fig:10vac}
\end{figure}

A PE contour on [010] plane of the $\rho_{vd}$ = 10\% is visualized using OVITO in its initial state at 1K (Figure \ref{fig:10vacin} ) and at 800K, the point of most negative TEC, in Figure \ref{fig:10vacfinal}.

The structure initially shows a stable lattice, but shows the formation of localized vacancy defect clusters that interact with each other triggering shortening effects within the system. As vacancy defects increase in concentration the localized cluster interactions cause macroscopic decreases to the TEC at an increasing rate. 

\begin{figure}
     \centering
     \begin{subfigure}[b]{0.4\textwidth}
         \centering
         \includegraphics[width=\textwidth]{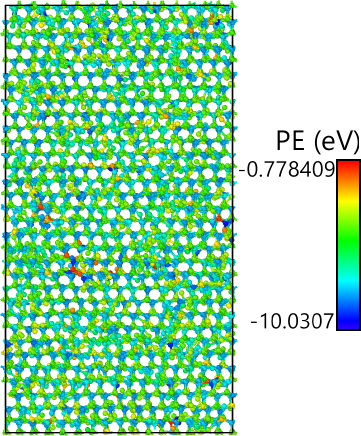}
         \caption{1K}
         \label{fig:10fpin}
     \end{subfigure}
     \hfill
     \begin{subfigure}[b]{0.4\textwidth}
         \centering
         \includegraphics[width=\textwidth]{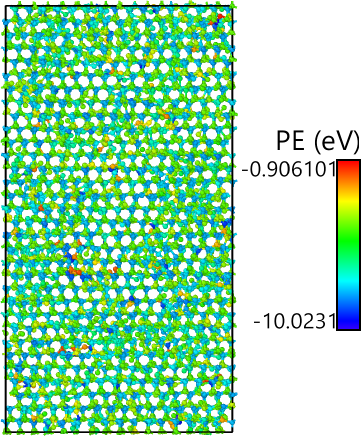}
         \caption{200K}
         \label{fig:10vfpfinal}
     \end{subfigure}
     \hfill
      \caption{a) Potential Energy contour of 10\% frenkel pair defected model at 1K and at  b) 200K }
    \label{fig:10vac}
\end{figure}

A PE contour on [010] plane of the $\rho_{fp}$ = 10\% is visualized using OVITO in its initial state at 1K (Figure \ref{fig:10fpin} ) and at 200K, the point of most negative TEC, in Figure \ref{fig:10vfpfinal}. The potential energy scale within the system remains during thermal equilibration. However, the frenkel pair defects interact with each other at this concentration to completely overwhelm the material by 200K. Where there are localized defect clusters at 1K, the material appears to be amorphous in regions in the system hardly keeping its lattice structure. The material at this concentration is not behaving like a solid lattice and suggests the material is prematurely changing states of matter. The results from the specific heat capacity study indicate that the energy required to raise the temerpature of the system is so low that the material sublimates.

\section{Conclusion}

We have shown the baseline structures have agreeing thermal properties with documented experimental and analytical values in literature. Observing defected 4H-SiC systems at various concentrations, Frenkel pairs show the highest magnitude difference per defect density in TEC and specific heat capacity in comparison with the pure structures. The variation in thermal expansion comes from the defects effect on the interatomic potential of the system. When defects are introduced the bonds and anaharmonicity in the system change resulting in changes in the internal energy. Changes in the internal energy consequently create deviation in the volume response via thermodynamic processes. We also document that defects at larger concentrations can cause NTE, which can cause severe incompatibility at material interfaces in electronic devices prompting degradation in the electrical performance. Though interstitials have the smallest effect on the thermal expansion, they have a much larger impact on effecting the internal energy of the system. Specific heat simulations show interstitials drop the energy needed to raise the temperature of the system over twice as much as the effect from the vacancies. The defect types showed decreases in $c_v$ at a decreasing rate suggesting the defects are interacting with each other to cause more rapid degeneration of the material. At high frenkel pair concentrations, the specific heat capacity reaches zero corresponding to an amorphous effect in the material and premature sublimation. This suggests the isothermal processes shift to adiabatic at this critical $\rho_{fp}$ Overall, the effects from displacement damage significantly alter the thermal properties of 4H-SiC as defect concentration increases. Alterations in the TEC and the specific heat can cause thermal stresses at material interfaces and quicker heating in power devices that can lead to permanent degradation or premature burnouts.

\clearpage

\bibliographystyle{ieeetr}
\bibliography{Bibliography}

\end{document}